\documentclass[onecolumn,preprint2]{aastex6}
\include{bibdef}

\def\et3{\eta_3}
\def\th1{\theta_{-1}}
\def\r07{r_{0,7}}
\def\x05{x_{0.5}}

\def\cm{\hbox{~cm}}

\def\Mpc{\hbox{~Mpc}}

\def\s{\hbox{~s}}

\def\MeV{\hbox{~MeV}}

\def\keV{\hbox{~keV}}

\def\G{\hbox{~G}}
\def\grb{{GRB~170817A}}

\def\erg{\hbox{~erg}}
\def\s{{\hbox{~s}}
\def\cm2{\hbox{~cm}^2}}

\def\Ep{E_{\rm peak}}

\def\Lobs{{1.6 \times 10^{47}~ \hbox{erg s}^{-1}}}
\def\r0{1.2 \times 10^{6}~ \hbox{cm}}
\def\es{\hbox{~erg s}^{-1}}

\newcommand{\appropto}{\mathrel{\vcenter{
  \offinterlineskip\halign{\hfil$##$\cr
      \propto\cr\noalign{\kern2pt}\sim\cr\noalign{\kern-2pt}}}}}

\begin{document}

\title{Gamma-ray burst models in light of the  GRB~170817A - GW170817 connection}

\author{P.~Veres}
\email{peter.veres@uah.edu}
\affiliation{Center for Space Plasma and Aeronomic Research, University of Alabama in Huntsville, 320 Sparkman Drive, Huntsville, AL 35899, USA}

\author{P.~M\'esz\'aros}
\affiliation{Department of Physics, The Pennsylvania State University, University Park, PA 16802, USA} 
\affiliation{Department of Astronomy \& Astrophysics, The Pennsylvania State University, University Park, PA 16802, USA} 
\affiliation{Center for Particle and Gravitational Astrophysics, The Pennsylvania State University, University Park, PA 16802, USA}

\author{A.~Goldstein}
\affiliation{Science and Technology Institute, Universities Space Research Association, Huntsville, AL 35805, USA}

\author{N.~Fraija}
\affiliation{nstituto de Astronom\' ia, Universidad Nacional Aut\'onoma de M\'exico, Circuito Exterior, C.U., A. Postal 70-264, 04510 M\'exico D.F., M\'exico.}

\author{V.~Connaughton}
\affiliation{Science and Technology Institute, Universities Space Research Association, Huntsville, AL 35805, USA}

\author{E.~Burns}
\affiliation{NASA Postdoctoral Program Fellow, Goddard Space Flight Center, Greenbelt, MD 20771, USA}

\author{R.~D.~Preece}
\affiliation{Space Science Department, University of Alabama in Huntsville, 320 Sparkman Drive, Huntsville, AL 35899, USA}
\affiliation{Center for Space Plasma and Aeronomic Research, University of Alabama in Huntsville, 320 Sparkman Drive, Huntsville, AL 35899, USA}

\author{R.~Hamburg}
\affiliation{Space Science Department, University of Alabama in Huntsville, 320 Sparkman Drive, Huntsville, AL 35899, USA}
\affiliation{Center for Space Plasma and Aeronomic Research, University of Alabama in Huntsville, 320 Sparkman Drive, Huntsville, AL 35899, USA}

\author{C.~A.~Wilson-Hodge}
\affiliation{Astrophysics Office, ST12, NASA/Marshall Space Flight Center, Huntsville, AL 35812, USA}

\author{M.~S.~Briggs}
\affiliation{Space Science Department, University of Alabama in Huntsville, 320 Sparkman Drive, Huntsville, AL 35899, USA}
\affiliation{Center for Space Plasma and Aeronomic Research, University of Alabama in Huntsville, 320 Sparkman Drive, Huntsville, AL 35899, USA}

\author{D.~Kocevski}
\affiliation{Astrophysics Office, ST12, NASA/Marshall Space Flight Center, Huntsville, AL 35812, USA}

\begin{abstract} For the first time, a short gamma-ray burst (GRB) was unambiguously associated with a gravitational wave (GW) observation from a binary neutron star (NS) merger. This allows us to link the details of the central engine properties to GRB emission models. We find that photospheric models (both dissipative and non-dissipative variants) have difficulties accounting for the observations. Internal shocks give the most natural account of the observed peak energy, viewing angle  and total energy. We also show that a simple external shock model can reproduce the observed GRB pulse with parameters consistent with those derived from the afterglow modeling. We find a simple cocoon shock breakout model is in mild tension with the observed spectral evolution, however it cannot be excluded based on gamma-ray data alone. Future joint observations of brighter GRBs will pose even tighter constraints on prompt emission models.
\end{abstract} 

\section{Introduction} \label{sec:intro} 
Gamma-ray bursts (GRBs) are intense flashes of gamma-rays and have been observed by space-borne observatories for decades \citep{klebesadel73}.  Recently, the discovery that binary neutron star mergers \citep{GBMLVC} result in short GRBs \citep{Goldstein,GBMLVC,MWL17170817a} closed the hunt for the progenitors of (at least some) short duration GRBs.

The first GW detection \citep{Abbott+16gw1} from binary black holes (BH) left open some intriguing possibilities that these types1 of GW mergers were followed by electromagnetic (EM) emission \citep{Connaughton+16gbmgw}. \citet{Zhang16gw,Perna+16gw,Loeb16gw} discussed different scenarios for extracting the required power in electromagnetic form from a binary BH merger. Assuming the BH binary was related to the observed short transient, GW150914-GBM, \citet{Veres+16gwgrb} discussed the implications of the joint detection for GRB models, however the weak nature of the gamma-ray signal did not allow for strong conclusions. Gamma-ray data for \grb, allows us to perform a more detailed analysis in the same vein.

It is not clear if \grb\, belongs to the classical version of gamma-ray bursts thought of as having a highly relativistic, narrow jet \citep{Goldstein} or if it has a Lorentz factor on the order of a few ($\lesssim 5$) and a more isotropic angular emission profile \citep{Kasliwal1559,Mooley+17cocoon}. This determination may come from ongoing afterglow observations \citep{Margutti+18170817aAG, Lyman+18170817aAG}. On the other hand we may need future observations of separate events to distinguish between these two models as they are both consistent with observations of \grb\, to date. Based on gamma-ray data, \grb, appears as a usual GRB with parameters that are within the observed range of previous GRBs.  
The low luminosity and the late detection of a rising afterglow are  strong indications that \grb~ was observed off-axis \citep{Troja+17170817a, 2017arXiv171008514F}.  
In addition, \cite{Salafia+16offaxis} points out that a lightcurve that is variable when seen on-axis, becomes smoother for larger viewing angles as it appears to be the case for \grb. 

It is still unclear what is the emission mechanism in the prompt phase of GRBs. A gravitational wave observation concurrent with a GRB detection gives new insights and constraints to this puzzle through information on the central engine.  The two competing models to explain the prompt emission are dissipative photosphere models \citep{Rees+05phot} and internal shocks \citep{Rees+94is}.  Internal shocks can account for the fast variability observed in some GRBs, while it seems to have problems explaining the preferred peak energy of an ensemble of GRBs \citep{Preece+00batse}, $\Ep\propto B \Gamma \gamma_e^2$, where $B$ is the local magnetic field, $\Gamma$ the bulk Lorentz factor and $\gamma_e$ is the random Lorentz factor of electrons with synchrotron peak at $\Ep$. A preferred $\Ep$ would require a fine tuning of three parameters which seems unlikely. The dissipative photosphere model accounts for the narrow distribution of peak energies, because it reflects the temperature of the innermost part of the jet that has a weak dependence on the available energy and the relevant radius has a restricted range: few times the Schwarzschild radius of a few Solar mass BH, $\Ep\propto L^{1/4}R_0^{-1/2}$. On the other hand, there are indications \citep[albeit relying on some assumptions, e.g.][]{Lyutikov06radius, Kumar+08radius} that point to an emission radius that is too large for the photospheric scenario.

In addition, external shock models have been considered in the past for explaining the prompt emission. External shocks develop, when the jet material is significantly slowed down by the circumstellar medium. This model is not unambiguously favored however, because some GRBs exhibit strong temporal variations, that could be difficult to explain by the interaction between the jet and the ISM \citep{Sari+97variab}. \citet{Dermer+99es} however argue, that the required variability can be achieved by the interaction of a blast wave with relatively small clouds with large density contrasts.

We thus consider it worthwhile to pursue tests of emission models for GRB prompt emission in the multimessenger era for gamma-ray bursts. See \citet{Begue+17170817a,Ming+18170817a} for different approaches on this topic.

The paper is arranged as follows. In Section \ref{sec:obs} we present the observations. In Section \ref{sec:unmod} we present model independent considerations, in Section \ref{sec:model} we present the relevant GRB models and apply them for \grb. We discuss our findings and present our conclusions and outlook in Section \ref{sec:disc}. We use the common notations for physical constants and present the scaling of quantities in the $Q_x=Q/10^x$ format in cgs units, unless otherwise noted.

\section{Observations} \label{sec:obs} 
\subsection{Gamma-rays} 
The lightcurve of \grb\, shows a relatively hard pulse of $\sim$0.5 s duration  followed by a weak soft emission approximately from T0+1 to T0+2 s, carrying about 1/3 of the total energy (T0 is the GBM trigger time). The start of the gamma-ray emission, defined as the first time the signal reaches 10\% of the peak rates, is $\Delta t_{\rm GRB-GW}=1.74\pm 0.05\s$ after the time of the BNS merger \citep{Goldstein}. The main pulse shows no obvious substructure, it is consistent with a single pulse. %

The spectrum of the main pulse  is best fit by a Comptonized model (power law function with a high energy exponential cutoff) having a photon index $=-0.62\pm 0.40$ and a peak energy $E_{\rm peak}=185 \pm 65 \keV$. The spectrum of the soft emission is consistent with a blackbody, with temperature $kT=10.3\pm1.5 \keV$ \citep{Goldstein} and with a normalization parameter equivalent to a radius $R_{\rm BB}=2.9 \times 10^8 (D_L/43\Mpc)$ cm. We use the luminosity distance  to the identified host galaxy of \grb, NGC 4993,  $D_L=43\pm 4 \Mpc\approx 1.32\times 10^{26} \cm$ reported in \citet{Coulter+17swope}.

The isotropic equivalent energy of the main pulse is $E_{\rm iso,Comp}=(4.0\pm1.0)\times 10^{46}$\,erg, and  $E_{\rm iso,BB}=(1.4\pm0.3)\times 10^{46}$\,erg for the soft pulse and the total energy released in gamma rays in the 1\,keV to 10 MeV range is $E_{\gamma,{\rm iso}}=(5.4\pm1.3)\times 10^{46}$\,erg. We derive a luminosity of $L_{\rm obs} = (1.6 \pm 0.6) \times 10^{47}$\,erg s$^{-1}$, in the 1\,keV--10\,MeV energy band \citep{GBMLVC}.  Observationally, \grb\, has ordinary properties in terms of duration, fluence or peak flux \citep{Goldstein}, however in context of the population with measured redshifts and hence luminosity, it is subluminous by orders of magnitude \citep{GBMLVC}.

\begin{figure*}
\centering
\includegraphics[width=0.4\textwidth]{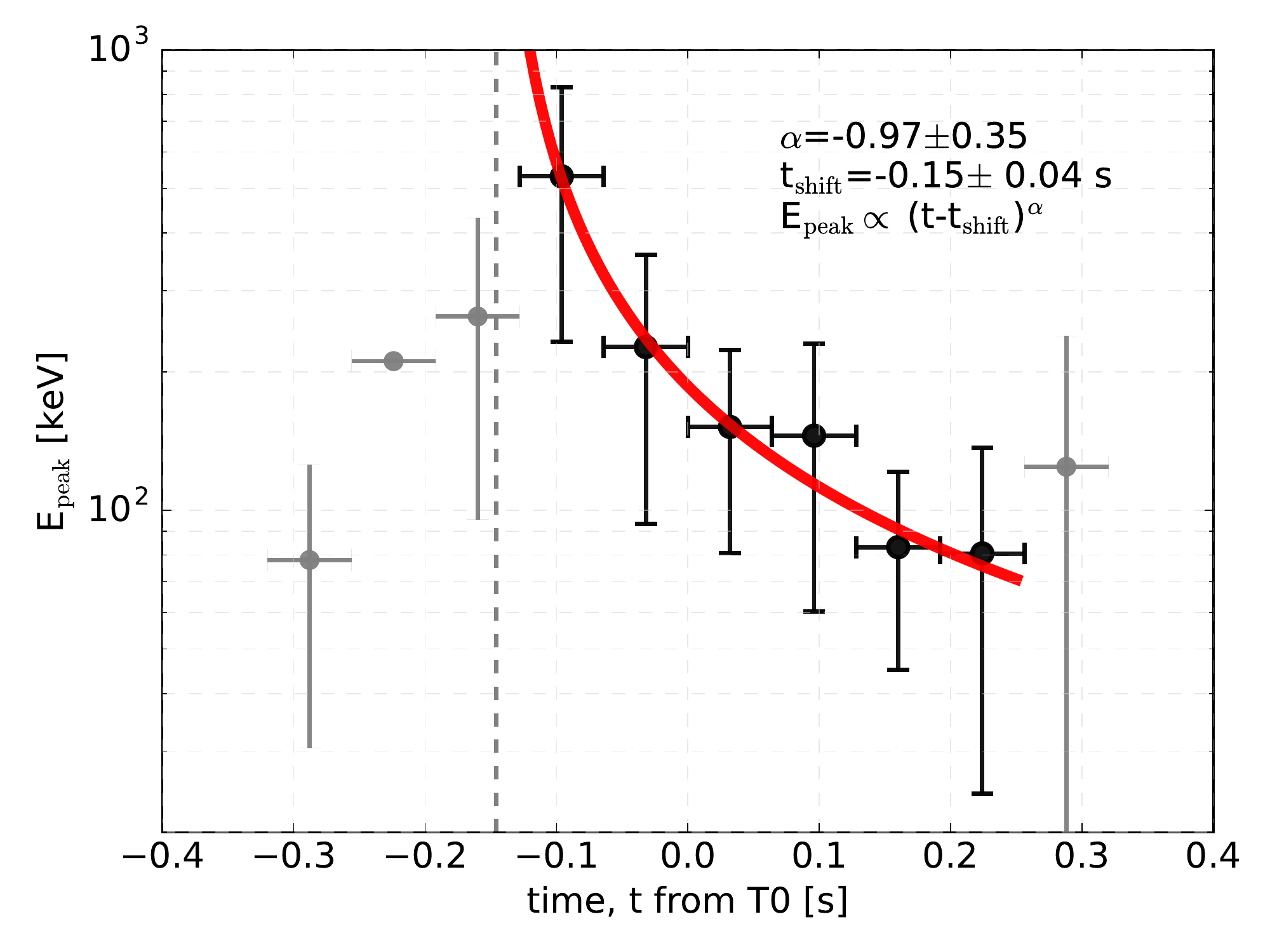}
\includegraphics[width=0.4\textwidth]{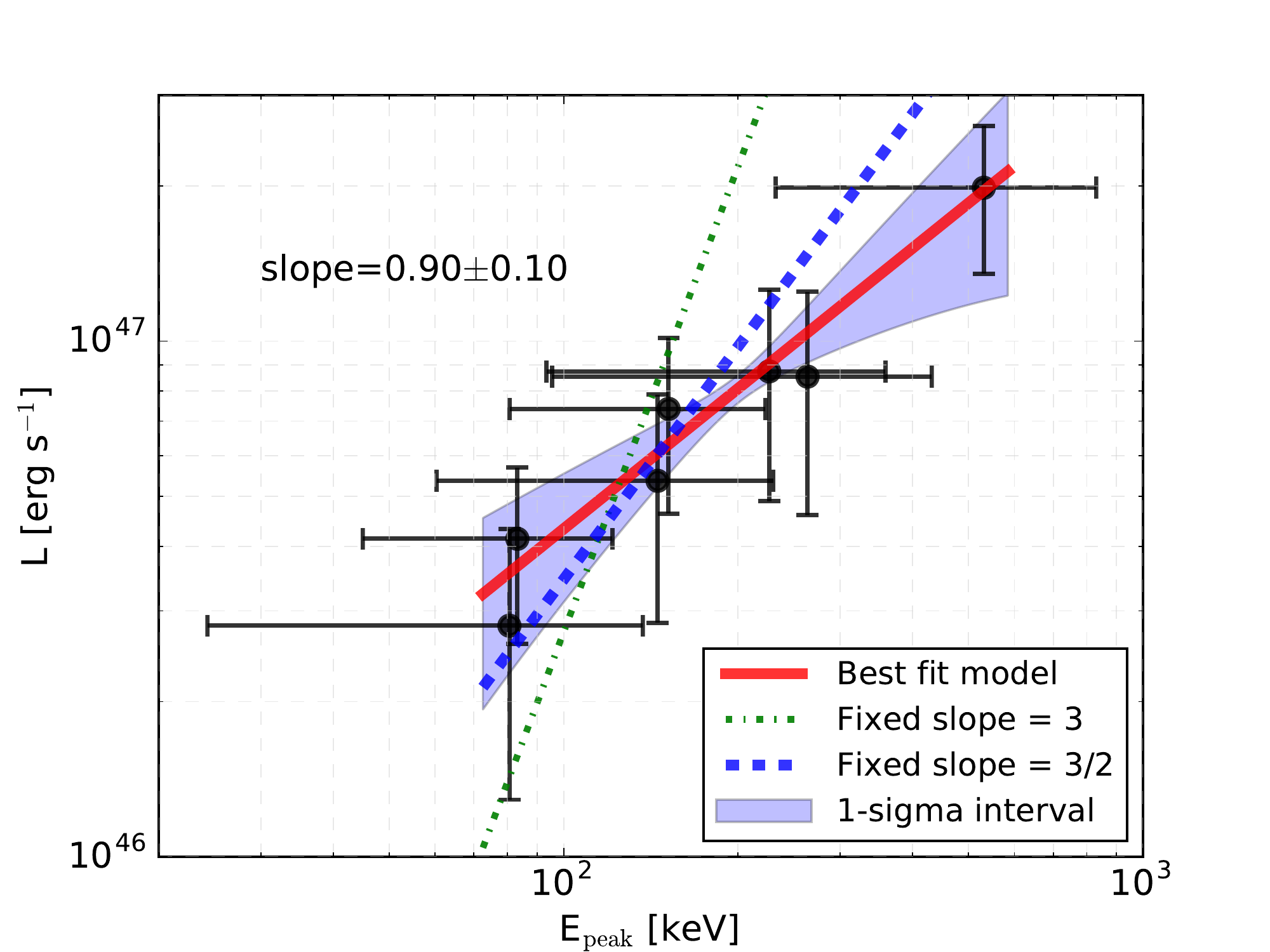}
\caption{Left: Peak energy evolution with time. We fit only the decaying phase. Right: Luminosity as a function of peak energy, $\Ep$. The red line is the overall best fitting power law. The green dash-dotted and blue dashed lines are power laws with indices {\it fixed} to 3 and 3/2 respectively, and illustrate cases for the briefly illuminated shell model (see Section \ref{sec:briefly}).}
\label{fig:eplum}
\end{figure*}

The hardness ratios indicate a hard to soft evolution with time during the main peak, characteristic of classical GRBs \citep{Hakkila+17h2s}.  Some models have particular, falsifiable predictions for the $\Ep(t)$ evolution \citep[e.g.][]{
1999ApJ...524L..47G,
dermer04, Genet+09-hilat, 
2012ApJ...751...33F, 
Preece+14130427agbm}.  To further investigate this trend, we have fitted the Comptonized model to gauge the evolution of the peak energy. In order to mitigate that the GRB is dim, we fix the photon index to the time integrated value and vary the amplitude and the peak energy, $\Ep$. Letting the photon index vary would result in similar $\Ep$ values but with larger uncertainties. GRBs with higher flux would allow for an unconstrained time resolved spectral fit.  Using this method we recover the hard-to-soft behavior (see Figure \ref{fig:eplum}, left) visible in the photon count data by fitting 7 bins (from T0-192 ms to T0+256 ms), with 64 ms resolution.  The decaying part of the time-$\Ep$ data yields a power law index of $\alpha=-0.97\pm 0.35$ and start time with respect to the GBM trigger time $t_{\rm shift}=-0.15\pm 0.04 \s$ ($E_{\rm peak} \propto (t-t_{\rm shift})^{\alpha}$). The temporal index, $\alpha$ strongly depends on the choice of the start time \citep[e.g.  see;][]{2017ApJ...848...94F, 2017ApJ...848...15F}. This is usually chosen as the start of the emission. If we fix the start time to $t_{\rm start}=T0-0.22 \s$ (the time the \citet{Norris+96pulse} pulse model reaches 5\% of the peak analogously to \citet{Goldstein}), we get $\alpha=-1.49\pm0.15$ which is consistent with the evolution of a synchrotron forward-shock  energy break in the fast-cooling regime.  The $\Ep$- luminosity data yields the following dependence:  $L\propto E_{\rm peak}^{0.90\pm 0.10}$ (see Figure \ref{fig:eplum}, right), and this does not depend strongly on the assumed fixed photon index.

The soft component (observed from T0+1 to T0+2 s) is clearly separated from the main pulse (T0-0.2 to T0+0.3 s), the hard-to-soft trend observed in the main pulse does not extend into the soft emission \citep[see also Figure 1 of][]{Pozanenko+17170817a}. We note that the time bin with the highest $\Ep(\approx 500\keV)$, is coincident with the  100 ms  peak for \grb, observed by INTEGRAL-ACS \citep{Savchenko+17grb170817a}. This is in line with ACS having a higher low-energy threshold than GBM.

\subsection{Gravitational waves} The advanced LIGO and Virgo observatories detected an event that is consistent with a binary neutron star merger \citep{Abbott+17BNS} with chirp mass of $M_c\approx1.188 M_\sun$.  Depending on prior assumptions on the spin of the neutron stars, the total mass can be $M_t\approx2.74 M_\sun$ (low-spin) or $M_t\approx2.82 M_\sun$ (high-spin) with somewhat larger uncertainty in the latter case.  For the purposes of our analysis, at the required accuracy we take the total mass to be $M_{t}=2.8 M_\sun$.

{\it Rotation parameter. } Because the merger occurs above the frequency range of LIGO and Virgo, no direct measurement of the final black hole rotation parameter is possible \citep{Abbott+17postmerger170817a}. From simulations \citep{Kiuchi+09rotpar}, the rotation parameter of a BH that forms within a dynamical timescale of the system and with mass ratio $q \gtrsim 0.8$ was found to be $a=0.78\pm0.02$.  The mass ratio does not cover all the allowed values by the LIGO/Virgo observations, however it is more characteristic of the low-spin case, based on observed neutron star spins. Thus for simplicity, we will take $a\approx0.8$ in our calculations. 

\section{Model independent considerations}
\label{sec:unmod}

\grb\, is unusually subluminous, it is located in the nearby universe and its spectrum does not extend above $\sim$1 MeV.  The low luminosity suggests either a high baryon load for the jet producing the GRB or a scenario where the jet is observed off-axis. Usual lower limits for characteristic Lorentz factors of $\eta\gtrsim100$ in order to escape the compactness problem \citep{goodman86} do not apply here, $\Gamma$ can be constrained to be $\gtrsim 2$ \citep{Kasliwal1559}. 

\subsection{\grb\, as an off-axis GRB}

An off-axis scenario naturally explains the low luminosity of \grb: we are observing the GRB jet from an angle to the jet symmetry axis ($\theta_v$) that is larger than the jet half opening angle ($\theta_j$). This picture is favored by long-wavelength observations \citep[e.g.][]{Troja+17170817a,Kasliwal1559}, however one needs to assume laterally structured jet and late energy injection into the afterglow shock\footnote{
Note that in the present paper we restrict ourselves to an investigation of the prompt emission of \grb.
The jet lateral structure in the late evolution phases leading to optical and radio emission may be different.}. 
It is also possible that the prompt emission was produced by the mildly relativistic shock breakout emission of a cocoon, which was energized by a jet (most likely pointing away from us, but also allowing for unsuccessful, choked jets) traveling through the merger debris \citep[e.g.][]{Kasliwal1559, Pozanenko+17170817a, Gottlieb+17cocoonprompt,Gottlieb+18cocoon, Hallinan+17170817a}.

In an off-axis scenario, it is difficult to ascertain the on-axis properties of the GRB (e.g. energetics, duration and spectrum) due to the strong dependence of the physical properties on the jet geometry and the unknown Lorentz factor (see however efforts by \citet{Troja+17170817a, Hallinan+17170817a, Margutti+18170817aAG}). In order to obtain simple scaling relations, we restrict ourselves to a top-hat jet, while in reality more complex jet profiles are expected. Indeed, late afterglow modeling \citep{Margutti+17170817a} suggests the lateral structure both in emissivity and Lorentz factor may be non-trivial, however the prompt emission occurs early after the merger and the top-hat jet model may be a good approximation \citep{Ioka+17170817a}.

The difference between quantities measured on- or off-axis can be expressed as powers of the ratio of Doppler factors  (with the caveat that e.g. isotropic energy calculations involving integrating emissivities over the surface of the jet involve careful calculations \citet{Ioka+17170817a}). Furthermore, it is convenient to assume that the viewing angle is not farther from the jet axis by more than twice the jet opening angle ($\theta_j<\theta_v<2 \theta_j)$. This assumption will simplify our discussion and it covers realistic scenarios for \grb. We define $b=1 + \Gamma^2 (\theta_v-\theta_j)^2$ to express the on-axis quantities as a function of the off-axis observations. The duration will be $T_{90}^{\rm off} \approx b T_{90}^{\rm on}  $, the peak energy $E_{\rm peak}^{\rm off} = b^{-1} E_{\rm peak}^{\rm on}$, the total energy $E^{\rm off} \approx b^{-2} E^{\rm on}$ and the luminosity: $L^{\rm off} \approx b^{-3} L^{\rm on}$.  Bursts with small viewing angles are necessarily brighter, more flux is expected above background, resulting in somewhat longer duration than from what is implied by the above scaling \citep{Zhang+06physproc}.

For the model-independent constraints we assume that the total isotropic equivalent energy ($E_{\rm iso}$), duration ($T_{90}$) and peak energy ($\Ep$) of \grb, if seen on-axis, would have been close to that of an average GRB (see Figure \ref{fig:offaxis}). For concreteness, we adopt the viewing angle to the jet axis (taken to be the binary NS axis of the angular momentum vector) to be $28^\circ$ \citep{GBMLVC}. For three  cases of the Lorentz factor (40, 100, 300) we calculate the duration, total and peak energy as a function of jet opening angle to illustrate what regions of the opening angle are consistent with the range of observed parameters in the Fermi GBM catalog \citep{Bhat+16cat}. The allowed parameter spaces  are shown in vertical, increasingly darker shades of red  regions in  Figure \ref{fig:offaxis}. The lesson from this analysis is that higher Lorentz factors require the jet to be viewed closer to the edge. However, even for reasonable viewing angles (e.g. $\theta_v-\theta_j\approx 10^{\circ}$), we find it would require \grb\, to have on-axis physical parameters that are extreme when compared to the observed distribution \citep[e.g. extremely short duration {\it and} extremely high peak energy][]{GBMLVC, Granot+17170817a}. The probabilities of observing \grb  with Lorentz factor of (40, 100, 300) are $P[\%]=(7.1,\,2.6,\,0.64)$ respectively.

\begin{figure*}
\centering
\includegraphics[width=0.8\textwidth]{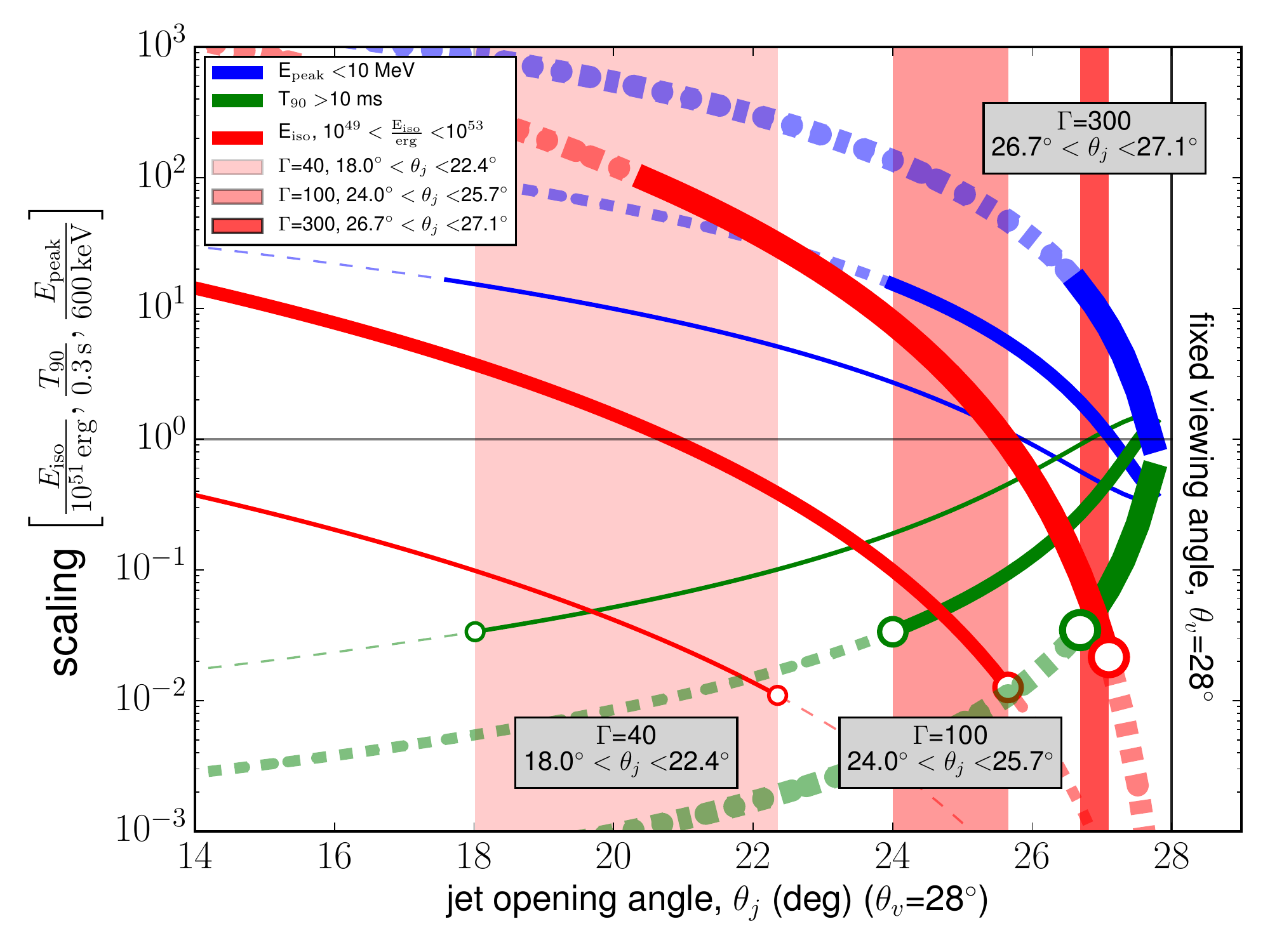}
\caption{Possible E$_{\rm peak}$ (blue), T$_{90}$ (green) and E$_{\rm iso}$ (red) measurements of \grb, for an {\it on-axis} observer. Values are scaled to average GRB measurements ($600 \keV$, {$0.3 \s$} and $10^{51} \erg$ respectively) and plotted as function of the jet opening angle. The viewing angle is fixed to $\theta_v=28^\circ$ \citep{GBMLVC}. Thin, medium and thick lines show Lorentz factors 40, 100 and 300 respectively. For each colored curve, the solid part marks values within the observed range for short GRBs (E$_{\rm peak}\lesssim10\MeV$, T$_{90}\gtrsim 10$\, ms and $10^{49} \erg\lesssim E_{\rm iso} \lesssim 10^{53} \erg$ ), the dashed portion indicates extreme values never before observed. Green and red circles mark the allowed ranges of opening angles for a given Lorentz factor  and these ranges are marked by shades of red. In these regions, E$_{\rm peak}$, T$_{90}$ and E$_{\rm iso}$ lie within a range of observed values for GRBs. Gray boxes indicate the range of allowed angles.
} \end{figure*}
\label{fig:offaxis}

\subsection{A briefly illuminated shell} 
\label{sec:briefly}
Irrespective of the detailed model for gamma-ray production, we can test basic physics assuming a relativistic shell emits radiation for a short time, then it is switched off. This approximation is closely related to how internal shocks operate.  Bearing in mind the large errors on the measured quantities ($\Ep$ and L), we can make the following observations.  The peak energy $\Ep$ is proportional to $t^{-1}$ which is a characteristic of a pulse that was illuminated for a finite duration, then switched off \citep{Genet+09-hilat,dermer04}. Just like in the case of the extremely bright GRB\,130427A however, the observed $L \propto \Ep$ behavior does not align with the expectations for this simple model. In the decaying phase of the pulse if the only contribution is from emission sites at increasingly higher latitudes, that would result in $L\propto \Ep^{3}$, instead of $L \propto \Ep$. Based on the green dot-dashed line on Figure \ref{fig:eplum}, right, we conclude that the index of 3 is not preferred. Invoking synchrotron emission, we have $\Ep\propto \Gamma B \gamma_e^2$ and $L\propto \Gamma^2 B^2 \gamma_e^2$, where $\gamma_e$ is the random Lorentz factor of the electrons and we assume $\Gamma=$constant in the coasting phase. Assuming adiabatic expansion for the emitting shell we have $\gamma_e\propto R^{-1}$. In the magnetic flux freezing regime, $B\propto R^{-2}$ and we recover $L \appropto \Ep^{3/2}$, that is marginally consistent with the data (blue dashed line on Figure \ref{fig:eplum}, right). The best fit to the data yields a slope of $\approx 1$ (red line on Figure \ref{fig:eplum}, right). If we parametrize the magnetic field as a function of radius as $B\propto R^{-l}$, we get $L\propto R^{-2l-2}\propto \Ep^{\frac{2l+2}{l+2}}$, and find that $l\approx0$ or B=constant provides the best solution.

\section{GRB models}
\label{sec:model}


We assume a BH formed after the BNS merger. A BH with a rotation parameter a=0.8, will have an innermost stable circular orbit of
\begin{equation}
R_{\rm ISCO}= R_0=f(a) R_G  = 2.91\frac{G M_{\rm BH}}{c^2} = 1.2\times 10^{6}~ (M_t/2.8 M_\sun) \cm,
\end{equation}
where f(a) is a function of the rotation parameter,
$f(a)=3+Z_2-[(3-Z_1)(3+Z_1+2Z_2)]^{1/2}$ and $Z_1=1+(2-a^2)^{1/3}[(1+a)^{1/3}
+ (1-a)^{1/3}] $ and $ Z_2=(3a^2+Z_1^2)^{1/2}$.
Around the newly formed black hole a large amount of energy is released in a small volume. The characteristic size of this volume will be approximately $R_{\rm ISCO}$. Subsequently a large entropy fireball is born, which accelerates until it reaches a relativistic bulk velocity \citep{Meszaros+93gasdyn,Piran+93hydro}.  The Lorentz factor evolves as $\Gamma=R/R_0$ up to the saturation radius defined by $R_{\rm sat}=\eta R_0=1.2\times 10^8~ \eta_2 (M_t/{2.8M_\sun}) \cm$.  At later times, the fireball starts decelerating as it interacts with the circumstellar material.

\subsection{Photospheric models}
In the relativistically expanding material the location where the Thompson
scattering optical depth falls below unity marks the location of the
photosphere.  This is the innermost radius from where radiation can escape and
can be calculated from $R_{\rm phot}= L_t\sigma_T/4\pi m_p c^3 \Gamma_{\rm
phot}^2 \eta$:
\begin{equation}
R_{\rm phot}\approx \left\{
\begin{array}{lll}
   R_0 \eta_T^{4/3} \eta^{1/3}= 1.1\times 10^{10} ~   \left(\frac{L_t}{\Lobs}\right)^{1/3} \left(\frac{\epsilon_\gamma}{0.2}\right)^{-1/3}
   \left(\frac{R_0}{\r0}\right)^{2/3} \eta_3^{-1/3} \cm
   	& {\rm if~ }  R_{\rm phot}<R_{\rm sat} & {\rm or~ } \eta>\eta_T\\
   R_0 \eta_T^4 \eta^{-3} = 9.4\times 10^{8} ~  \left(\frac{L_\gamma}{\Lobs}\right)\left(\frac{\epsilon_\gamma}{0.2}\right)^{-1} \eta_2^{-3} \cm
   	& {\rm if~ }  R_{\rm phot}>R_{\rm sat}& {\rm or~ } \eta<\eta_T,
	\end{array}
	\right. 
\end{equation}
where the observed gamma-ray luminosity ($L_\gamma$) is a fraction
$\epsilon_\gamma$ (efficiency) of the total luminosity ($L_t$) and
\begin{equation}
\eta_T=\left(\frac{ L_t \sigma_T}{4\pi m_p c^3 R_0}\right)^{1/4}\approx 167~\left(\frac{L_\gamma}{\Lobs }\right)^{1/4} \left(\frac{\epsilon_\gamma}{0.2}\right)^{-1/4} \left(\frac{R_0}{\r0}\right)^{-1/4}
\label{eq:etat}
\end{equation}
is the Lorentz factor separating the photosphere in the coasting phase ($\eta<\eta_T$) and the photosphere in the acceleration phase ($\eta>\eta_T$). $\Gamma_{\rm phot}$ is the Lorentz factor at the photosphere.

\subsubsection{Non-dissipative photosphere models}
\label{sec:ndis}
 The temperature of an
expanding fireball at its base can be calculated as
\begin{equation}
T_0\approx (L_t /4\pi R_0^2 c a)^{1/4}\approx 321 (L_\gamma/ \Lobs )^{1/4} (\epsilon_\gamma/0.2)^{-1/4}
(R_0/\r0)^{-1/2}\keV. 
\label{eq:T0}
\end{equation}

{\it Photosphere in the acceleration region -\,} In case the photosphere occurs in the acceleration region, we observe the same temperature as the initial fireball. This is because of the linear increase of the Lorentz factor ($\Gamma\propto R$) counteracts the comoving temperature dropping as $T'\propto R^{-1}$ and the observed temperature scales approximately as $T=T' \Gamma\sim$ constant.  

Even though the spectrum is non-thermal, it is possible to achieve a broadened spectrum solely from photospheric photons by including scattering and geometrical effects \citep[e.g.][]{peer05}.  The observed peak corresponds to $\Ep=185 \keV \equiv 3.92 kT_0$, where the numerical factor comes from the peak of the Planck spectrum in $\nu F_\nu$ representation. Substituting into Equation \ref{eq:T0} we get $(\epsilon_\gamma/0.2) (R_0/\r0)^2=2140$. By requiring the efficiency, $\epsilon_\gamma<1$ we get the estimate for the launching radius $R_0 \gtrsim  60 R_G= 2.5\times 10^7 \cm$. The limiting Lorentz factor in this case will be $\eta_T\approx 52$, and the above scenario can be easily accomplished by a coasting Lorentz factor $\eta>\eta_T$.  $R_0\gtrsim 60 R_G$ suggests that the start of the acceleration does not occur in the immediate vicinity of the newly formed BH.  We note that in this case, the observed delay between the launch of the jet and the photosphere is on the order $\Delta t_{\rm delay, phot.} = 10^{-5} (L_\gamma/\Lobs)^{-1/3} \epsilon_{\gamma,0}^{1/3} (R_0/2.5\times 10^7\cm)^{4/3} \eta_2^{-1/3} \s$, meaning the observed 1.74 s delay between the merger and the onset of the GRB does not pose a useful constraint and can be ascribed to other effects.

{\it Photosphere in the coasting phase -\,}
In the coasting phase, the observed temperature will decrease
with radius as $T=T_0 (R/R_{\rm sat})^{-2/3}$. The condition for $R_{\rm ph}>R_{\rm sat}=\eta R_0$ in terms of critical Lorentz factors can be stated as
$\eta<\eta_T\approx167$ (see equation \ref{eq:etat}). In other words, low
Lorentz factors, $\lesssim$ 100 will result in the photosphere occurring in the
coasting phase of the jet.

The peak will occur at
\begin{equation}
E_{\rm pk}^{\rm PH}= 3.92 \times kT_0 \left(\frac{R_{\rm phot}}{R_{\rm sat}}\right)^{-2/3} =E_{\rm pk}^{\rm obs}\approx 185 \keV
\label{eq:phpksat}
\end{equation}
and we get $R_{\rm phot}\approx 2.3 R_{\rm sat}=2.7\times 10^{8} (L_\gamma/\Lobs)^{3/8} (\epsilon_\gamma/0.2)^{-3/8} (R_0/\r0)^{1/4} \eta_2 \cm$. This radius does not give a delay that is comparable to the delay observed between the GW and GRB signal. We may ask what parameters are needed in the non-dissipative photosphere scenario so that the GW-GRB delay is accounted for by the jet propagation time, $\Delta t_{\rm GRB-GW}=1.74 \s$. We can write $R_{\rm ph}= R_0 \eta_T^4 \eta^{-3} = 2 c \Delta t_{\rm GRB-GW} \eta^2 $ and have 
\begin{equation}
\eta=R_0^{1/5} \eta_T^{4/5} (2c\Delta t_{\rm GRB-GW} )^{-1/5} = 6.2
~(L_\gamma/\Lobs)^{1/5} (\epsilon_\gamma/0.2)^{-1/5}.
\label{eq:etaphsat}
\end{equation}
Note that the last equation is independent of the launching radius, and indeed points to a low Lorentz factor for which the photosphere occurs in the coasting phase. Furthermore, such a low Lorentz factor is equivalent to the approximately isotropic emission we expect from a cocoon shock-breakout event.

{\it Viewing angle effects -\,}
In this section we introduce the effects of viewing angle, and test the photospheric model's consistency with data. For an on-axis jet, the {\it maximum} peak energy will be given by Equation \ref{eq:T0}. If we approximate the relation between the on- and off-axis isotropic-equivalent luminosity as $L_{\rm t}^{\rm on}\approx L_{\rm t}^{\rm off} b^3$, where $b=1+\Gamma^2(\theta_v-\theta_j)^2$, we have $T_{0,{\rm max}}\propto b^{3/4}$, and: 
\[E_{\rm peak}^{\rm on}=b E_{\rm peak}^{\rm obs} \lesssim 321 ~b^{3/4} \keV. \] 
This is an upper limit of the peak energy of the non-dissipative photosphere model. Substituting the observed values, we have \[\theta_v-\theta_j\lesssim1.7 ^\circ~ \eta_2^{-1} (L_\gamma/\Lobs)^{1/2} (\epsilon_\gamma/0.2)^{-1/2} (R_0/\r0)^{-1}. \]

The afterglow observations \citep{Margutti+17170817a,Hallinan+17170817a} find an off-axis viewing angle $\theta_v-\theta_j \gtrsim 5^\circ$. The above limit together with the multiwavelength observations suggests that the non-dissipative photosphere may have some problems explaining the observed peak energy, given the luminosity and launching radius. We note that the launching radius provided by the GW observations provides is a minimum value that in this equation translates to a maximum $\theta_v-\theta_j$ angle.  This reasoning is similar to the one put forward by \citet{Fan+12corr, Zhang+12epeak, Veres+12peak, Veres+16gwgrb} but generalized for arbitrary viewing angle. Thus we conclude that in a top hat jet scenario the observations of \grb\, suggest that the non-dissipative photosphere model cannot be at work.

\subsection{Dissipative photosphere models}
\label{sec:dissphot}

Dissipative photosphere models involve energy released below or close to the photosphere \citep{Rees+05phot}. Variants include neutron-proton collisional heating \citep{Beloborodov10phot} or magnetic reconnection \citep{Giannios12peak}. The latter case might even involve a different acceleration region compared to the case when most of the energy in the jet is carried by baryons. Instead of $\Gamma\propto R$ we can have $\Gamma\propto R^{1/3}$ or more  generally $\Gamma\propto R^{\mu}$, $1/3\lesssim\mu\lesssim 1$ \citep{Giannios+10grbmag,McKinney+11switch,Meszaros+11gevmag,Bosnjak+12delay,Veres+12magnetic}.  For ease of calculation we can estimate the properties of the dissipative photosphere as arising from synchrotron emission of shocked electrons close to the photosphere of the outflow \citep{Meszaros+11gevmag, Veres+12magnetic, Veres+12fit}. 

The photosphere will occur at $R_{\rm phot}=R_0 \eta_T^{1/\mu} (\eta_T/\eta)^{1/1+2\mu}$ if it is in the acceleration phase ($\eta>\eta_T$) or R$_{\rm phot}=R_0 \eta_T^{1/\mu} (\eta_T/\eta)^{3}$ if in the coasting phase ($\eta<\eta_T$). $\eta_T=(L_t\sigma_T/4\pi R_0^2 m_p c^3 )^{\mu/(1+3\mu)}$ is the limiting typical Lorentz factor obtained from equating the photospheric radius with the location where the Lorentz factor saturates.  For simplicity we neglect the pairs created around the photosphere. This would have the effect of increasing the photospheric radius by up to one order of magnitude \citep{Meszaros+01pair,Veres+12magnetic}. The peak of the spectrum  will occur at $E_{\rm peak}=\gamma_e^2\Gamma_{\rm phot} \frac{q_e B}{2\pi m_e c}$, where B is the magnetic field, assumed to carry some fraction $\epsilon_B\lesssim 1$ of the total energy density, $B\approx (32 \pi n'_b m_p c^2 \epsilon_B \Gamma_r^2)^{1/2}$, and $n'_b=L/4\pi R_{\rm phot}^2 m_p c^3 \eta \Gamma_{\rm phot}$ is the comoving baryon number density, $\Gamma_r\gtrsim 1$ is the relative Lorentz factor of the shells participating in semirelativistic shocks, and $\gamma_e\approx m_p/m_e\Gamma_r$ is the typical random Lorentz factor of the shocked electrons.

We examine if this model is able to reproduce the observed peak energy with the assumption that we are viewing the jet off-axis, in a top-hat jet scenario.  We start from the observed and total luminosity as in section \ref{sec:ndis}, assuming it resulted from an off-axis observation. The actual value would be $L_{\rm t}^{\rm ON}\approx b^3 L_{\rm t}^{\rm OFF}$. We calculate the peak energy in a dissipative photosphere model using  $L_{\rm t}^{\rm ON}$ to obtain $E_{\rm peak}^{\rm ON}$, the on-axis value. Keeping in mind that the observed peak energy, $E_{\rm peak}^{\rm obs} =b^{-1} E_{\rm peak}^{\rm ON}$ we solve for $E_{\rm peak}^{\rm obs}\approx 185 \keV$.

\begin{figure*}
\centering
\includegraphics[width=0.49\textwidth]{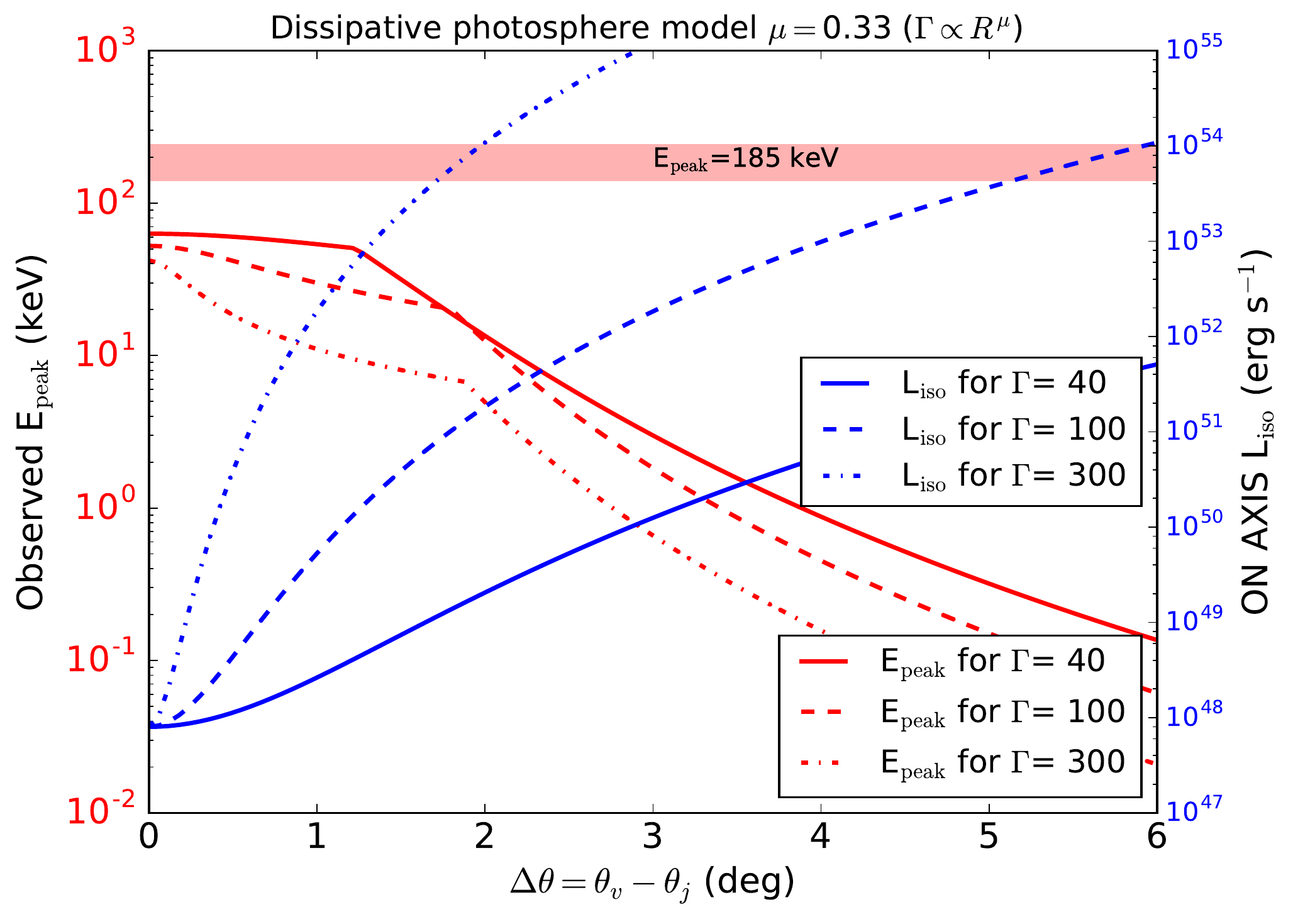}
\includegraphics[width=0.49\textwidth]{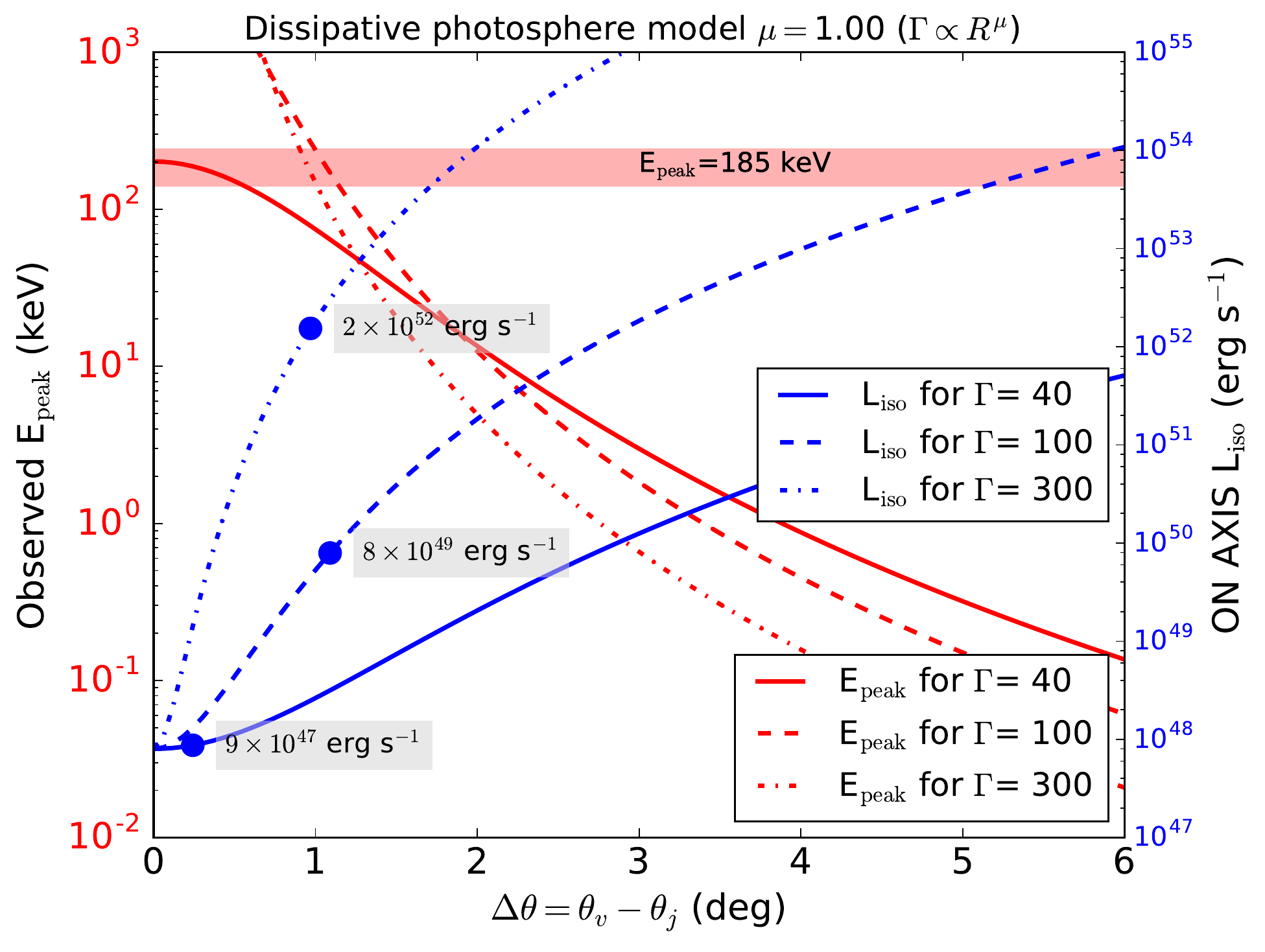}
\caption{Observed peak energy and total, on-axis luminosity as a function of the off-axis angle for three different Lorentz factor values. In the magnetic field dominated case (left, $\mu=1/3$), the observed peak energy is impossible to recover even for the most favorable set of parameters. In the baryon dominated case (right, $\mu=1$), the observed peak energy can be obtained (where the red curves intersect the horizontal $\Ep=185\keV$ line) with a range of luminosities for different Lorentz factors (blue dots), all suggestive of viewing angles close to the jet edge, $\Delta \theta\sim 1^{\circ}$. }
\label{fig:dissphot}
\end{figure*}

We find that in the magnetically dominated dissipative photosphere model ($\mu=1/3$, $\Gamma\propto R^{\mu}$), it is impossible to reach the observed peak energy for even the most favorable set of parameters (see Figure \ref{fig:dissphot}, left). We find a solution if the relative Lorentz factor of the shocked shells is in excess of $\Gamma_r \gtrsim 10$, but we consider that to be too extreme. 

In the baryon dominated case ($\mu=1$), $E_{\rm peak} \approx 185 \keV$ can be obtained for the three representative Lorentz factors (40, 100, 300), and results in total luminosities in the range $9\times 10^{47}$ to $2\times 10^{52}$ erg s$^{-1}$. These scenarios arise for off-jet angles ($\Delta \theta=\theta_v-\theta_j$) of order $\sim 1^\circ$ (see Figure \ref{fig:dissphot},  right). Such small angles, similarly to the non-dissipative models, pose problems  for the dissipative photosphere scenario.

\subsection{Internal shocks} The main pulse is consistent with a single emission episode. In the internal shock scenario \citep{Rees+94is} an emission episode corresponds to the collision of two shells ejected at different times. Internal shocks will occur in the coasting phase of the jet, because collisions are improbable in the acceleration phase \citep[e.g.][]{Ioka+11Lorentz}.  Another limiting radius   is the external shocks radius \citep{lee05} that marks the limit of efficient energy extraction before the jet energy is tapped by the interstellar medium. 

{\it Dynamical considerations -\,} We calculate in  detail the collision of two equal mass shells \citep{Kobayashi+01is,daigne98,barraud05}.  We mark the time of the binary NS merger by $t_0$. The launching of the first shell occurs at $t_1=t_0+\Delta t_{\rm wait}$ with Lorentz factor $\Gamma_1$. The second shell is launched at $t_2=t_1+\Delta t_{\rm launch}$ with $\Gamma_2=\kappa \Gamma_1$, $\kappa>1$. We condense the lifetime of the hypermassive NS until it collapses to a BH and other effects local to the central engine into $\Delta t_{\rm wait}$. This timescale will be the same in the lab frame and the observer frame, neglecting cosmological effects.  The two shells collide at $t_{\rm IS}=t_1+\Delta t_{\rm launch} (1-\kappa^{-2})^{-1}$ and at the radius $R_{\rm IS}=2 c \Delta t_{\rm launch} \Gamma_m^2 (\kappa-\kappa^{-1})^{-1}$ \citep[e.g.][]{Krimm+07IS}.  In the $\Gamma_{1,2}\gg 1$ limit $\beta=\sqrt{1-1/\Gamma^2}\approx 1-1/2\Gamma^2$ and $\Gamma_m=\sqrt{\Gamma_1 \Gamma_2}=\Gamma_1 \sqrt{\kappa}$ is the Lorentz factor of the merged shell in the approximation where the shells have equal mass.

The efficiency of internal shocks is the amount of energy that is available to be radiated away compared to the kinetic energy of the shells $\eta_{\rm eff} =1-2( \kappa^{1/2}+ \kappa^{-1/2})^{-1}$. Typical efficiencies for single collisions are in the range of a few percent \citep{Kobayashi+97variab}.

If we are viewing the jet at an angle larger than the jet opening angle ($\Delta \theta=\theta_v-\theta_j>0$), an additional delay arises from the geometry  $\Delta t_{\rm geom}= R_{\rm IS} (1-\cos \Delta\theta)/c$ that sets constraining upper limits on the emission radius  \citep{Granot+17170817a}.

The observed delay between the launch of the first shell (activation of the central engine) and the start of the gamma-ray emission will be \citep{Fenimore+99IS,Krimm+07IS}:
\[
\Delta t_{\rm IS}= t_{\rm IS}-R_{\rm IS}/c=t_1+\frac{\Delta t_{\rm launch} }{1-\kappa^{-2}}
\]

The total time budget of the delay between the GWs and the start of the GRB, $\Delta t_{GRB-GW}=1.74 \s$ must come from these three sources:
\[
\Delta t_{\rm wait} + \Delta t_{\rm IS} + \Delta t_{\rm geom} =1.74 \s
\]

\begin{figure}
\centering
\includegraphics[width=0.99\textwidth]{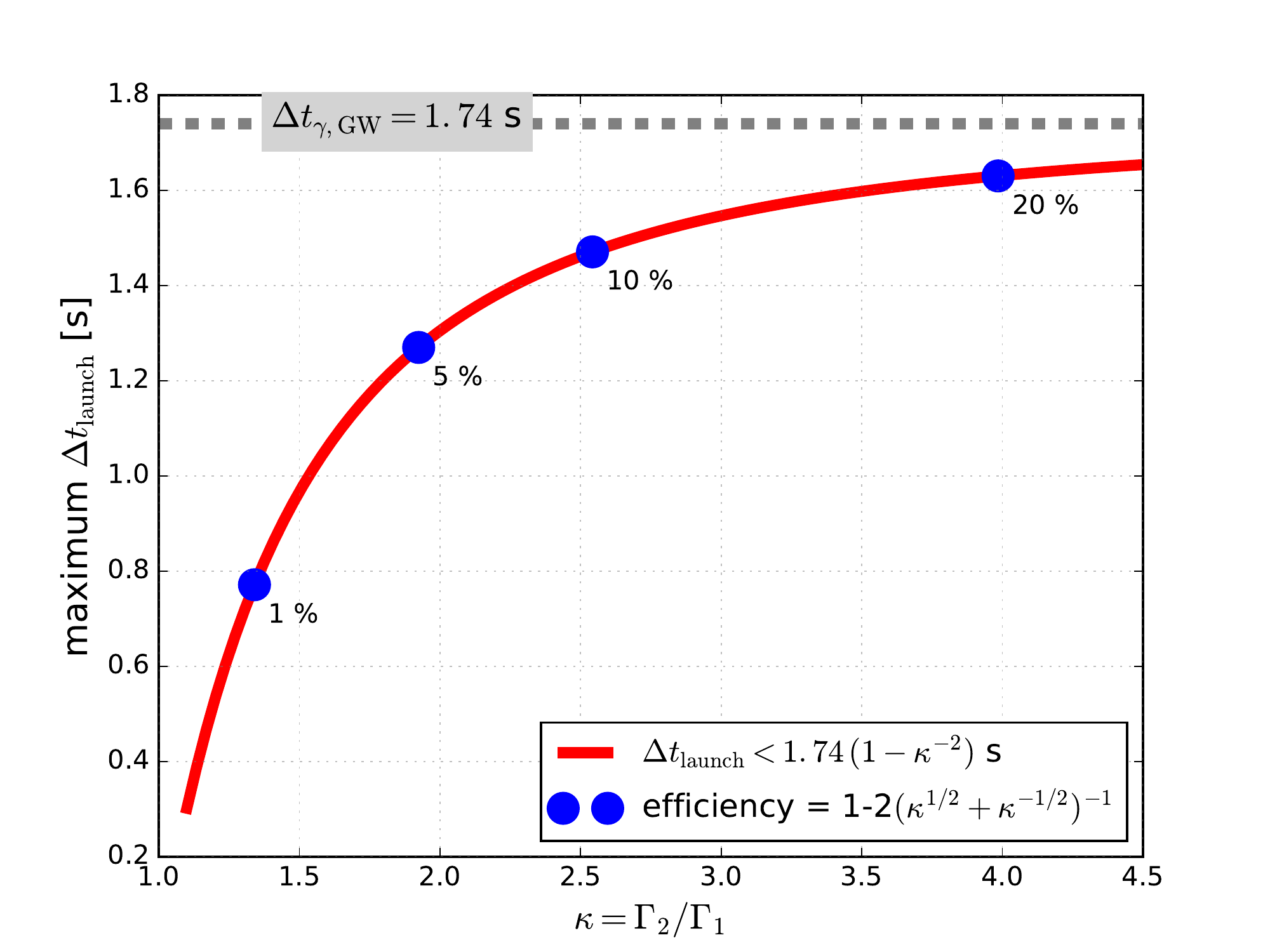}
\caption{Maximum delay between the launching of the two shells as a function of the Lorentz factor contrast so as the entire delay between the gravitational waves and gamma-rays is ascribed to this delay. Numbers next to the blue dots mark the efficiency of the collision. }
\label{fig:dtIS} \end{figure}

In the case, where the time delay is exclusively accounted for by producing the internal shocks, we can put upper limits on the time the central engine was active ($\Delta t_{\rm launch}$, see Figure \ref{fig:dtIS}). For example, assuming the spread of Lorentz factors ejected by the central engine is on the order of the individual Lorentz factors or $\Delta \Gamma\approx \Gamma$, implying $\kappa\approx 2$, for a single collision we have $\approx 5\%$ efficiency of converting bulk kinetic energy to internal energy, and the central engine can be active at most $\sim 1.3 \s$ ($\sim$70 \% of the total delay). For lower efficiency of 1 \%, we have $\Delta t_{\rm launch}<0.8 \s$ ($\sim$40 \% of the total delay). Similar limits can be placed on the emission  radius,
\[
R_{\rm IS}<2c\Delta t_{\rm obs} \Gamma_m \frac{1-\kappa^{-2}}{\kappa -\kappa^{-1}}
\]
however, these limits range from few times $10^{13} \cm $ for $\Gamma_m=40$ to a few times $10^{15} \cm$ for $\Gamma=300$ with a weak dependence on the Lorentz factor contrast, $\kappa$. These limits are within the expected range of emission radii for internal shocks and are not constraining.

{\it Spectral considerations -\,} The internal shock radius is often determined using the variability timescale, which for this GRB is $dt_{\rm var} \approx 0.1 \s$ \citep{Goldstein}. We have $R_{\rm IS}\approx 2c t_{\rm var} \eta^2 = 6 \times 10^{13} (dt_{\rm var} /0.1 \s) \eta^2_2$  cm.  Gamma-rays are produced by synchrotron radiation from shocked electrons and the characteristic Lorentz factor of the shocked electrons can be expressed as $\gamma_m=\epsilon_e/\zeta_e m_p/m_e (p-2)/(p-1) \Gamma_{\rm r}$, where $\Gamma_{\rm r}$ is the relative Lorentz factor of the shells and $\zeta_e$ is the fraction of electrons that are shock accelerated. The comoving particle density in the shocked region is $n'= L / 4 \pi R_{\rm IS}^2 m_p c^3 \eta^2$.  We adopt the scenario where the magnetic fields are built up by the shock up to some $\epsilon_B$ fraction of the equipartition energy \citep{medvedev99}.  The magnetic field strength is $B=(8\pi\epsilon_B \Gamma_{r} n' m_p c^2)^{1/2}=5.4\times 10^5 (L_\gamma/1.6\times 10^{47} \es)^{1/2} \epsilon_{B,-1}^{1/2}  \epsilon_{e,-1}^{-1/2} \eta^{-3}_1 dt_{{\rm var},-1}^{-1} \Gamma_{r,0} \G$.

The synchrotron peak energy will be \citep[e.g.][]{lee05}: 
\[
h \nu_m=\Gamma \gamma_m^2  \frac{q_e B}{m_e c} = 0.13~  (L_\gamma/1.6\times
10^{47} \es)^{1/2} \epsilon_{B,-1}^{1/2}  \epsilon_{e,-1}^{3/2} \eta^{-2}_1
(dt_{\rm var}/0.1 \s)^{-1} \zeta_{e,-1}^{-2}  \Gamma_{r,0}^3 \keV. 
\]

\begin{figure}
\centering
\includegraphics[width=0.99\textwidth]{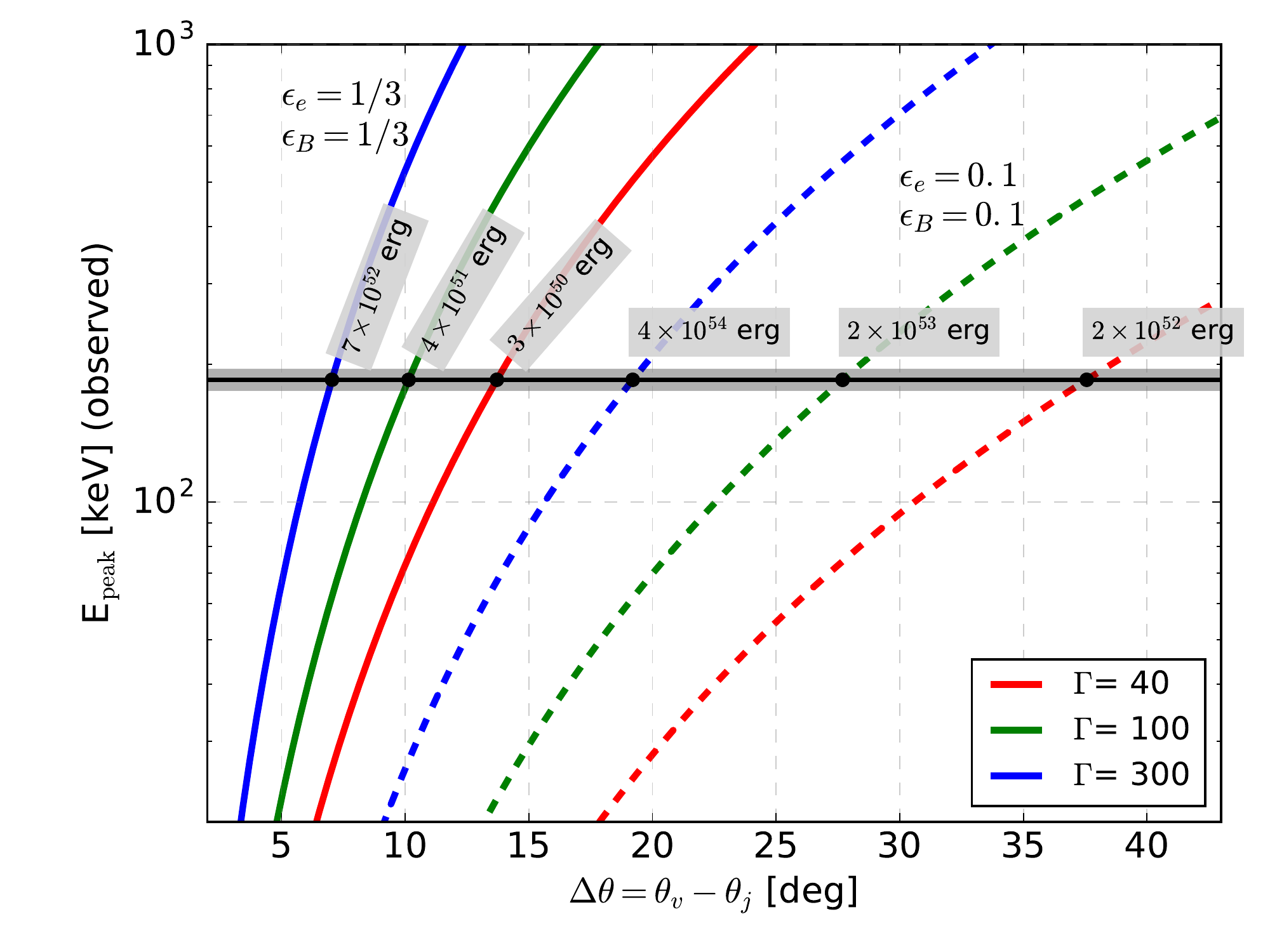}
\caption{Curves showing the peak energy as a function of off-axis angle for three
different Lorentz factor values in the internal shock scenario. The two groups of curves  are for two different
choices of microphysical parameters (solid: $\epsilon_e=\epsilon_B=1/3$, dashed: $\epsilon_e=\epsilon_B=0.1$). The gray boxes indicate the total energy
for an {\it on-axis} observer for the different solutions.} \label{fig:EpIS}
\end{figure}

Introducing viewing angle effects in a similar way to Section \ref{sec:dissphot}, we are asking if there is a plausible set of parameters which can explain the observed peak energy for a given off-axis angle and bulk Lorentz factor. We find that for $\epsilon_e=\epsilon_B=1/3$ (solid lines in Figure \ref{fig:EpIS}), reasonable total energy ($3\times 10^{50}$ to $7\times 10^{52} \erg$), within the observed range, can explain $\Ep=185 \keV$ for off-axis angles of order 10 degrees. A different choice of microphysical parameters ($\epsilon_e=\epsilon_B=0.1$) yields a larger set of angles and total energy for the three representative Lorentz factors (40, 100, 300).

\subsection{External shocks}
\label{sec:es}
External forward shocks \citep{Meszaros+97ag, Sari+98ag} are invoked usually to model the multiwavelength afterglow. The prompt emission of simple (smooth lightcurve) GRBs can be also modeled by external shocks \citep{Panaitescu+98promptES,Dermer+99es}

The surrounding region of the BNS merger is expected to be of constant density and very tenuous with particle densities similar to the intergalactic medium, $n\sim 10^{-5}-10^{-3} \cm^{-3}$. However BNS modeling efforts indicate that in the last stages of the merger material can be ejected from the system covering the direction of the rotation axis, making an environment similar to a wind medium. For these reasons we test both the constant density and the wind medium version of the external shocks. In order to show that external forward shocks are a viable scenario, we follow  \citet{Dermer+99es} and \citet{dermer99}  with some modifications to create a gamma-ray lightcurve.

\begin{figure}
\centering
\includegraphics[width=0.4\textwidth]{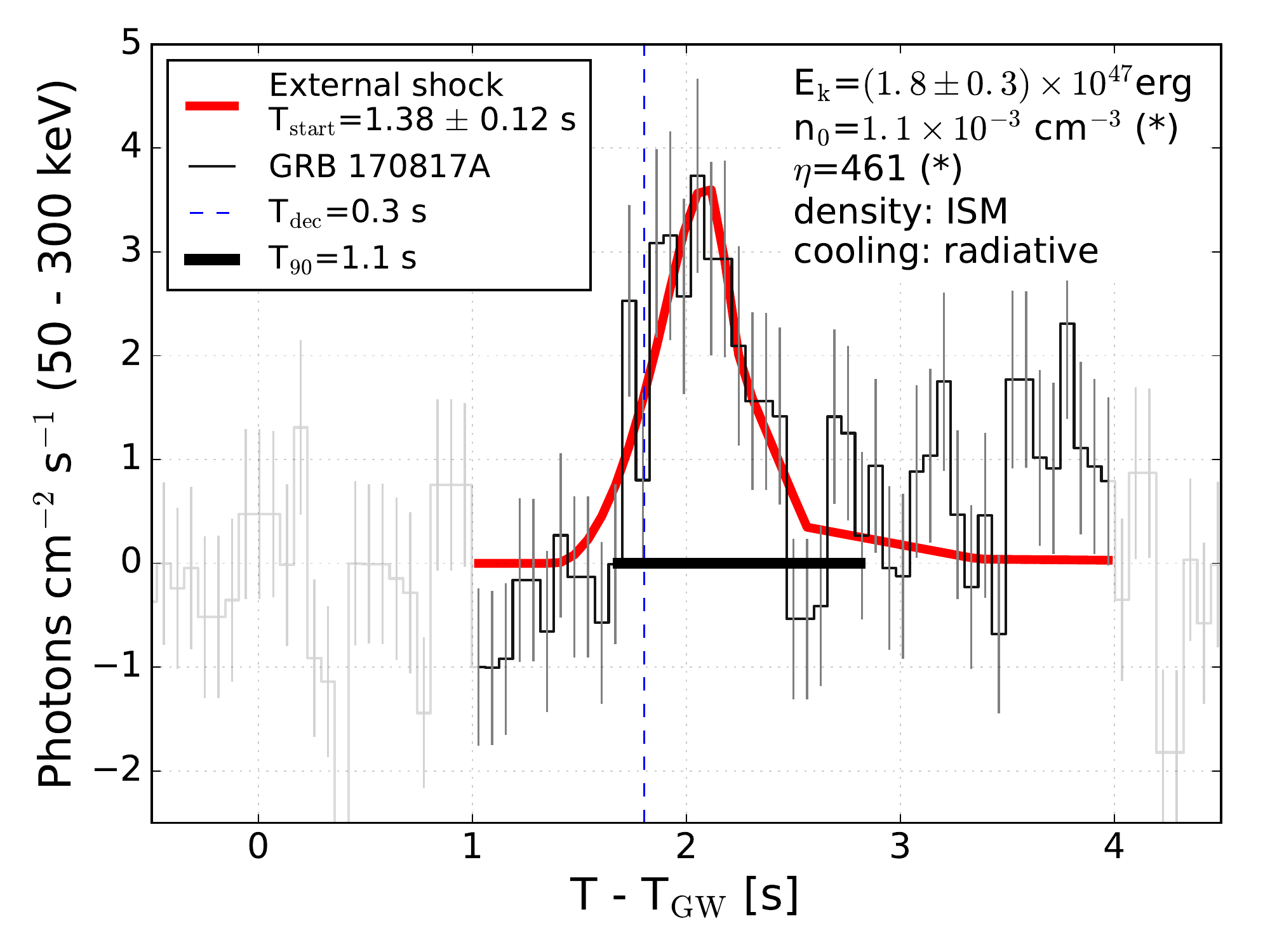}
\includegraphics[width=0.4\textwidth]{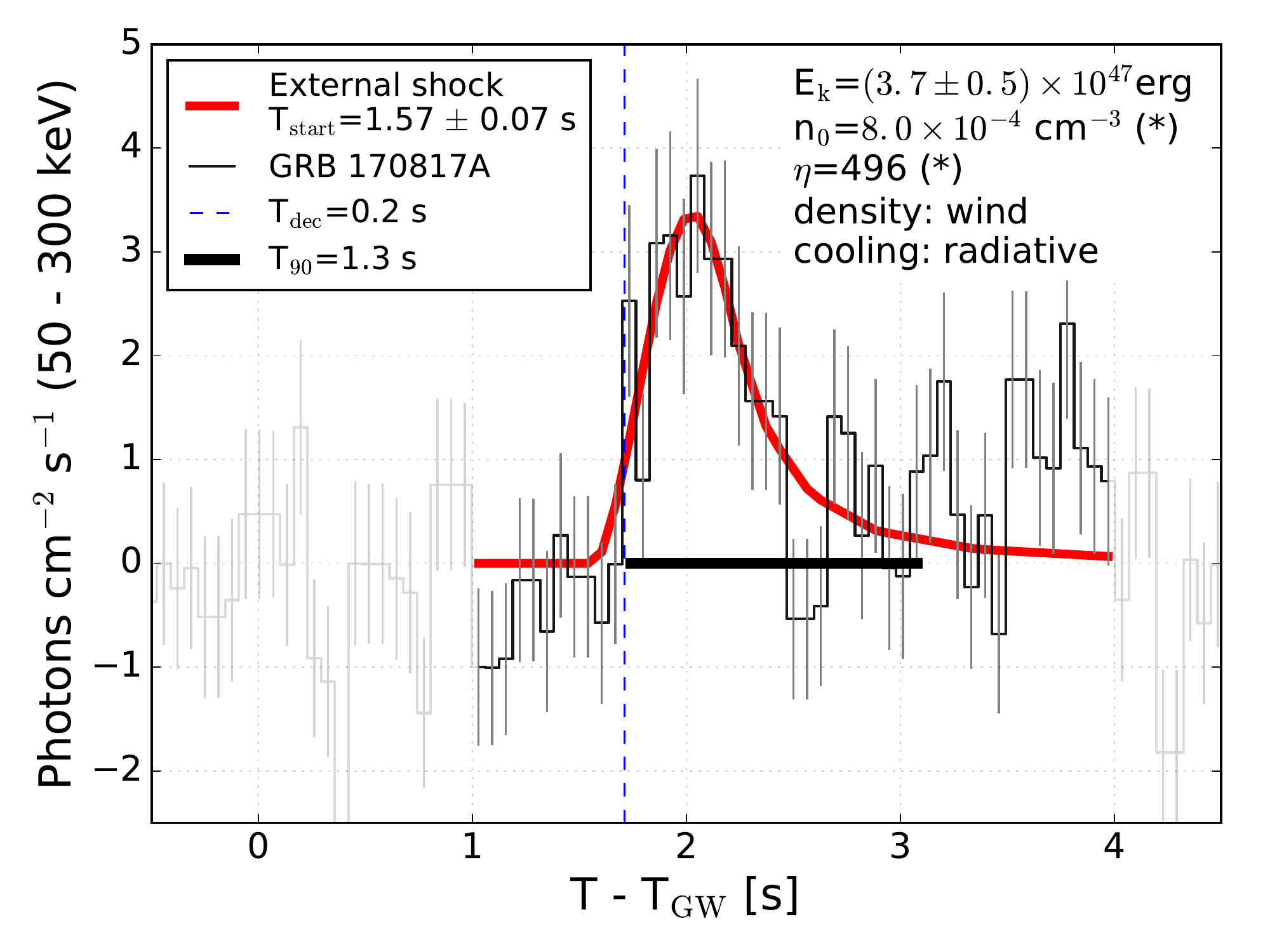}
\caption{External shock model, reproduces the observed pulse for the constant density model (left) or in wind medium (right). The start time and the total kinetic energy can be constrained. The values for the Lorentz factor and the density at the deceleration are degenerate, indicated by (*), however the density is close to the actual value (see Section \ref{sec:es}). The dashed blue vertical line indicates the deceleration time, the thick black line shows $T_{90}$ for the model lightcurve and as expected they are comparable. Heavy data lines mark the time interval where the data and model are compared.}
\label{fig:lcES}
\end{figure}

In the external shock model both the peak time and the duration are governed by the deceleration time. When only restricted data is available, we can assume the deceleration time is equal to the duration, however in concrete case like \grb,  more elaborate calculations can be performed. We indeed show that both of these quantities are on the order of the deceleration time (see horizontal solid black  and vertical dashed blue lines on Figure \ref{fig:lcES}).   

The deceleration radius, $R_d$, is defined in a medium with density $n=n_0 (R/R_d)^{-k}$ as the distance up to which $1/\eta$ fraction of the initial fireball's energy is plowed up from the interstellar medium. $R_d=((3-k)E_k/4\pi n_0 m_p c^2 \eta^2)^{1/3}$ and $n_0$ is the density at the deceleration radius. This approach for calculating the deceleration radius is consistent with the more widely used method  \citep[see e.g.][]{chevalier99}, keeping in mind that here $n_0$ can be a function of the deceleration radius (e.g. in the wind case).

The Lorentz factor as a function of radius is constant up to $R_d$, asymptotically reaching $R^{-g}$ where $g$ is in the range from 3/2 to 3 depending on the radiative regime (radiative or adiabatic) and interstellar density profile (constant density or wind). To have a smooth transition we use the functional form $\Gamma(R)=\eta \left(1+(R/R_d)^{g s}\right)^{-1/s}$, with the smoothness parameter, $s$, fixed arbitrarily to 3.

The time evolution of the flux is described by a broken power law at every instant: 
\[P(E,t)=\frac{(1+\alpha_{\rm high}/\alpha_{\rm low}) P_p(t)}{(E/E_{\rm peak}(t))^{-\alpha_{\rm low}} + \alpha_{\rm high}/\alpha_{\rm low} (E/E_{\rm peak}(t))^{-\alpha_{\rm high}}},\] 
where $\alpha_{\rm low}$ ($\alpha_{\rm high}$) is the slope of the $\nu F_\nu$ spectrum below (above) the peak. $P_p(t)\propto (\Gamma(R)/\eta)^4 (R/R_d)^{2-k}$ is the amplitude of the $\nu F_\nu$ peak and $E_p(t)=E_{\rm {peak},0}(\Gamma(R)/\eta)^4 (R/R_d)^{k/2}$ describes the temporal evolution of the peak energy. To recover the observed lightcurve from the model above, we integrate between 50 and 300 keV and set $E_{\rm {peak},0}$ so that the time averaged peak energy will be $\sim 200$ keV as observed. 

We use the GBM {\tt bcat}\footnote{\url{https://heasarc.gsfc.nasa.gov/FTP/fermi/data/gbm/triggers/2017/bn170817529/current/glg_bcat_all_bn170817529_v01.fit}} data product (typically used to determine the duration) to compare the external shock model with the observations.  Each time-bin is separately fitted by a power law function with an exponential cutoff and thus the final product is a flux curve, suitable for comparison with the model. 

The external shock model gives a remarkably good fit to this single pulse GRB both in the ISM and in the wind case (see Figure \ref{fig:lcES}). The model we fit has 4 free parameters: start time ($t_{\rm start}$), total kinetic energy ($E_k$), density at the deceleration radius ($n_0$) and coasting Lorentz factor ($\eta$). The start time and total kinetic energy could be constrained, while the density and the Lorentz factor have a degeneracy which prevents us from constraining them. Examining the goodness of fit distribution however on the $n_0-\eta$ plane, we find that along a 'valley' described by  $n_0 \approx 10^{-3} (\eta/500)^{-1/8} \cm^{-3}$, the resulting fits are indistinguishable. This indicates that while the Lorentz factor can't be constrained, the recovered density is close to the actual value within a factor of few.
For the constant density, ISM medium we find 
$t_{\rm start}-T_{\rm GW}=1.38 \pm 0.12 \s$,
$E_k=(1.8\pm 0.3 \times 10^{47} \erg$ and
$n_0=1.1\times 10^{-3} \cm^{-3}$.
For the wind medium we have 
$t_{\rm start}-T_{\rm GW}=1.57 \pm 0.07 \s$,
$E_k=(3.7\pm 0.5) \times 10^{47} \erg$ and
$n_0=8.0\times 10^{-4} \cm^{-3}$. It is remarkable that the density from afterglow modeling using sophisticated jet profiles, $n\approx 10^{-4} \cm ^{-3}$ \citep[e.g.][]{Margutti+18170817aAG} is within a factor of 3 of the value from the simple external shock modeling of the prompt emission. Such modeling is useful as it yields the start time of the emission based on a physical model instead of an arbitrary functional form for the pulse.

\section{Discussion and conclusion}
\label{sec:disc}
We have analyzed the prompt emission of the first gamma-ray burst unambiguously associated with a binary neutron star merger. First, we have adopted model independent constraints, based on the observed quantities of Fermi-GBM GRBs to point out the possible geometries within the top-hat jet model. Next we considered a briefly illuminated shell model (a proxy for internal shocks) that is a framework for explaining the spectral evolution in the prompt phase.  We have tested both the dissipative and non-dissipative versions of the photospheric models and found them unlikely to be at work in the case of \grb. We also found that the internal shock model can explain the observed spectra in the most straightforward manner. We tested external shocks as well, and found it also plausible as it reproduces the lightcurve with reasonable parameters.

Based on the range of observed GRB parameters in the Fermi GBM sample, we find, solely on probabilistic grounds, that the off-axis, top-hat jet model points to a relatively low Lorentz factor jet. Indeed, for lower Lorentz factors, the allowed region (seen in Figure \ref{fig:offaxis}) covers a wider range of angles which translates to a higher probability of observing such an event. On the other hand, for the allowed range of opening angles the on axis parameters for \grb\, are extreme: low total energy, very short duration and high peak energy. This points to the shortcomings of the top-hat jet model if \grb\, is indeed part of the highly relativistic, classical GRB population.

In the briefly illuminated shell model we found that the temporal evolution of the peak energy follows the simple but robust prediction of the model, a power law of index $\approx$-1. Notably this fit also yields a start time of the emission which is in agreement with the observed start of gamma-ray photon emission. For future joint detections, especially for brighter bursts, this represents an important tool to infer the launching time of the relativistic jet. Such a measurement will put limits on the lifetime of the hypermassive neutron star, and provide the launch time of the jet.

In the internal shock scenario, we have put limits on the time the central engine was active. While the limits are not strongly constraining, it provides a window into the duration of the central engine activity, which can be expanded on with future observations. When allowing for off-axis viewing angles, the internal shock model can naturally account for the observed peak energy with realistic total energy and viewing angle.

The external shock model gives a reasonable fit to the deconvolved flux lightcurve. Surprisingly, the inferred density in the ISM case is consistent with those derived from afterglow modeling even though the modeling uses re-energized shocks. We find that external shock modeling can constrain the kinetic energy,  start time of gamma-ray emission and the external density.

As an alternative to the relativistic jet scenario, the cocoon model explains the prompt phase as shock breakout from a cocoon that has moderate Lorentz factor \citep{Gottlieb+17cocoonprompt}. It is important to note that the cocoon scenario also involves a jet (either successful or unsuccessful in piercing through the NS merger debris), that through interaction with the merger debris, creates and energizes a cocoon. Testing against the observations would require cocoon simulations which is beyond the scope of this work; however, investigating  simple scaling relations can be informative. In cocoon models the matter in the outflow is usually assumed to have a power law velocity distribution, parametrized as $E(>\beta\gamma)\propto(\beta\gamma)^{-s}$, where $s$ is a free parameter. Analytic considerations and simulations both indicate \citep[e.g.][]{Tan+01grbsn,Kasliwal1559} values of $s\approx5$. E.g. \cite{Wang+18cocoon,Piro+17cocoon} derive scaling relations for a cocoon breakout model. We assume that the peak energy evolution described in Section \ref{sec:obs} corresponds to the evolution of the temperature in the cocoon model.  According to \citet[e.g.][]{Wang+18cocoon}, we have $kT\propto (t-t_{\rm start})^{-\frac{1}{2} - \frac{1}{8-2s}}$ and $L\propto kT^{\frac{12}{5+s}}$  while the measured power law indices found from fitting are approximately $kT\propto (t-t_{\rm start})^{-1}$ and $L\propto kT$. The $L-kT$ evolution is approximately reproduced by the observations for $s \gtrsim 7$. The peak energy (or equivalently, the temperature) temporal index in the cocoon scenario for $s \gtrsim 7$ is $\gtrsim -0.6$, while we measure it around $-1.0$. By shifting the start time of the pulse the fit can mimic lower indices, closer to $-1$ and consistent with the model. To achieve this, however the start time, $t_{\rm start}$ has to be shifted to a time with already significant emission (e.g. a temporal index of -0.5 would require $t_{\rm shift}-T0=-0.10\pm0.01 \s$ and the emission starts at $\approx T0-0.3\s$). We thus find some tension for the observed spectral evolution within the cocoon model; however the large errors on the time resolved data and the simplistic modeling makes a stronger conclusion unjustified. Future comparison of a stronger GRB and a detailed cocoon model will test the cocoon model further.

In conclusion, we find that the measurements  of the prompt emission favor the internal shock model, since it can reproduce the observations with a realistic set of parameters. We disfavor the photospheric models as they indicate viewing angles very close to the edge of the jet. We also find that the external shock model is a promising alternative to the above models.

Future joint observations will provide additional delay measurements between GWs and gamma-rays. The fine structure of the lightcurve will reveal if there is indeed  shorter timescale variability, on top or instead of the currently observed simple structure of the pulse. Such variability will provide a launching radius constraint which can be compared to the GW measurements of the central engine.

P.V. thanks Dr. M. Barkov for discussions and  acknowledges Fermi grant NNM11AA01A, Fermi GI grant 80NSSC17K0750 and partial support from OTKA NN 111016. P.M. acknowledges partial support from NASA NNX13AH50G.

\bibliography{pvall,grb}
\bibliographystyle{apj}

\end{document}